\documentclass{ws-ijmpe}
\usepackage[super]{cite}
\usepackage{graphicx}
\usepackage[colorlinks,citecolor=blue,bookmarks]{hyperref}
\usepackage{lscape}

\begin{document}
\markboth{B. Kumar et al.}{Shape co-existence....}
\catchline{}{}{}{}{}

\title{Shape co-existence and parity doublet in Zr isotopes} 
\author{\footnotesize Bharat Kumar\footnote{Email: bharat@iopb.res.in},
S. K. Singh, S. K. Patra}
\address{Institute of Physics, Bhubaneswar-751 005, India.}

\maketitle

\begin{history}
\received{(received date)}
\revised{(revised date)}
\accepted{(Day Month Year)}
\end{history}

\begin{abstract}
 We studied the ground and excited states properties for Zr isotopes starting
from proton to neutron drip-lines using the relativistic and non-relativistic
 mean field formalisms with BCS and Bogoliubov pairing. 
The celebrity  NL3 and SLy4 parameter sets are used in the calculations.
 We find spherical ground and low-lying
largedeformed excited states in most of the isotopes. Several couples of
$\Omega^{\pi}=1/2^{\pm}$ parity doublets configurations are found,
while analyzing the single-particle energy levels of the largedeformed 
configurations. 
\end{abstract}
\ccode{PACS Number(s): 21.10.Dr,21.10.Hw,21.10.Ft,21.10.Pc}

\section{Introduction}\label{introduction}
Although the nuclear shape co-existence for various
mass regions of the periodic table is a well known phenomena,
it remains an interesting investigation till today. On the other hand,
the existence of parity doublet  is relatively new~\cite{singh14,arima93}.
The origin and manifestation of such an interesting observable
is not yet known clearly. It is reported that the parity doublet is
not visible in a nucleus with normal/spherical deformation.
However, the existence of parity doublet is possible for nuclei 
with highly deformed shape. In this case, two orbitals
with opposite parity lie very close to each other.
Since, the parity doublet is only appeared in largedeformed configuration and
not in normal or spherical shape, the possibility of its origin may be
related to its shape, i.e. with deformed orbitals. That means, in normal 
situation, the high lying
partner of the doublet does not come nearer to the low lying one but 
when the nucleus gets deformed, gives rise a Nilsson like structure in the
largedeformed state.
The shape co-existence, i.e., two different shapes
with very close in energy is also a rare, but known incident in
nuclear structure physics \cite{rana,raj,sara,egido,shaib}. 
In this case, both the solutions are nearly or  completely
degenerate (different configuration with same energy).
This phenomenon is mostly visible in the mass region $A=100$ 
of the periodic table\cite{petrovici12}. Here, we have chosen Zr nucleus 
as a potential candidate both for shape co-existence and study of parity doublets using the 
well known relativistic (RMF) and non-relativistic (SHF) mean field formalisms.
The NL3 and SLy4 parametrization with BCS and Bogoliubov pair
prescriptions used to take care of the pairing for the open shell
nuclei.\\
The paper is organized as follows: In sections~\ref{shf} and~\ref{rmf}, 
we have given a brief outline about the non-relativistic 
Skyrme-Hartree-Fock-Bogoliubov (SHFB) and relativistic mean field (RMF) 
formalisms. 
Our results are discussed in section~\ref{result}. 
A concluding remark is given in section~\ref{remark}. 


\section{Skyrme-Hartree-Fock-Bogoliubov Approximation :}\label{shf}
The energy density functional with Skyrme-Hartree-Fock-Bogoliubov Approximation
is a powerful theoretical formalism to deal with finite nuclei starting from
both proton to neutron drip-lines\cite{{dean03}}. 
In this calculations, we have used the most successful SLy4 parameter 
set~\cite{chabanat98} with zero-range Bogoliubov pairing interaction for 
open shell nuclei. 
The numerical calculations are done using an axially deformed Harmonic 
oscillator (HO) basis state expansion to solve the Schr\"odinger equation 
iteratively. 
The numerical calculations are carried out using the code HFBTHO 
(v1.66p)~\cite{ring05} that solve the equation self-consistently.
For Skyrme forces, the HFB energy has the form of a local energy density 
functional~\cite{ring80,gam97,Per04,ben03}:
\begin{equation}
E[\rho,\tilde{\rho}]=\int d^3{\bf r}~{\cal H}({\bf r}) , \label{shfb}
\end{equation}
where, Hamiltonian density ${\cal H}$:
\begin{equation}
{\cal H}({\bf r})=H({\bf r})+\tilde{H}({\bf r})
\label{enden}
\end{equation}
is the sum of the mean-field and pairing energy densities. 
In the present implementation, we use the following explicit forms:
\begin{equation}
\begin{array}{rll}
H({\bf r}) & = & \tfrac{\hbar^{2}}{2m}\tau +\tfrac{1}{2}t_{0}\left[~
\left( 1+\tfrac{1}{2}x_{0}\right) \rho^{2}\right. -\left(
\tfrac{1}{2}+~x_{0}\right) \sum\limits_{q}\left.
\rho_{q}^{2}~ \right] \\
& + & \tfrac{1}{2}t_{1}\left[~ \left( 1+\tfrac{1}{2}x_{1}\right)
\rho \left( \tau -\tfrac{3}{4}\left. \Delta \rho \right) \right.
\right] -\left(\tfrac{1}{2}+~x_{1}\right) \sum\limits_{q}\left.
\rho
_{q}\left( \tau_{q}-\tfrac{3}{4}\Delta \rho_{q}\right) \right] \\
& + & \tfrac{1}{2}t_{2}\left[~ \left( 1+\tfrac{1}{2}x_{2}\right)
\rho \left( \tau +\tfrac{1}{4}\Delta \rho \right) \right. -\left(
\tfrac{1}{2}+~x_{2}\right) \sum\limits_{q}\left. \rho
_{q}\left( \tau_{q}+\tfrac{1}{4}\Delta \rho_{q}\right) \right] \\
& + & \tfrac{1}{12}t_{3}\rho^{\alpha }\left[ \left( 1+\tfrac{1}{2}
x_{3}\right) \rho^{2}\right. -\left( x_{3}+\tfrac{1}{2}\right)
\sum\limits_{q}\left. \rho_{q}^{2}\right] \\
& - & \tfrac{1}{8}\left( t_{1}x_{1}+t_{2}x_{2}\right)
\sum\limits_{ij}{\bf J} _{ij}^{2}   +  \tfrac{1}{8}\left(
t_{1}-t_{2}\right) \sum\limits_{q,ij}{\bf J}
_{q,ij}^{2} \\
& - & \tfrac{1}{2}W_{0}\sum\limits_{ijk}\varepsilon_{ijk}\left[
\rho {\bf \nabla }_{k}{\bf J}_{ij}\right. +\sum\limits_{q}\left.
\rho_{q}{\bf \nabla }_{k}{\bf J}_{q,ij}\right], 
\label{skyrmeph}
\end{array}
\end{equation}
\begin{equation}
\displaystyle \tilde{H}({\bf r})=\tfrac{1}{2}V_0
\left[1-V_1\left(\frac{\rho}{\rho_0}\right)^\gamma~
\right]\sum\limits_{q}\tilde{\rho}_{q}^{2}. 
\label{edhppd}
\end{equation}
The index $q$ labels the neutron ($q=n$) or proton
($q=p$) densities, while densities without index $q$ denote
the sums of proton and neutron densities.
$H({\bf r})$ and $\tilde{H}({\bf r})$ depend on the
particle local density $\rho ({\bf r})$, pairing local density
$\tilde{\rho}({\bf r})$, kinetic energy density $ \tau ({\bf r})$,  
and spin-current density ${\bf J}_{ij}({\bf r})$. 
The number of oscillator shells $N_{sh}=20$ to avoid the convergence problem 
and basis parameter $b_0=\sqrt{b_z^2+b_\bot^2}$ are used in the calculations. 
A detail numerical technique is available in Ref.~\cite{ring05}
and the notations are their usual meaning.

\subsection{Pairing Correlations in SHF formalism}\label{lnp}
In non-relativistic Skyrme-Hartree-Fock-Bogoliubov (SHFB) formalism,
we have included pairing correlation by using Lipkin-Nogami (LN)
prescription~\cite{ring05,lipkin60}.
In this calculation, the LN method is implemented by perturbing the SHFB 
calculation with an additional term $h^{'}=h-2\lambda_{2}(1-2\rho)$ is 
included in the HF Hamiltonian, where the parameter $\lambda_2$ is iteratively 
calculated so as to properly describe the curvature of the total energy as a 
function of the particle number. 
For an arbitrary two-body interaction $\hat{V}$, $\lambda_2$ can be 
calculated from the particle number dispersion according the following 
relation~\cite{ring05}:
\begin{eqnarray}
\lambda_2=\frac{<0|\hat{V}|4><4|\hat{N^2}|0>}{<0|\hat{N^2}|4><4|\hat{N^2}|0>},
\end{eqnarray}
where $|0>$ is the quasiparticle vacuum, $\hat{N}$ is the particle number 
operator, and $|4><4|$ is the projection operator onto the 4-quasiparticle 
operator space. 
The final expression for the $\lambda_2$ can be written in following simple 
form~\cite{flocard97}:
\begin{eqnarray}\label{ll2}
\lambda_2=\frac{1}{2}\frac{\text{Tr}\Gamma^{\prime}\rho(1-\rho) 
+ \text{Tr}\Delta^{\prime}(1-\rho)\kappa}{[\text{Tr}\rho(1-\rho)]^2 
-2 \text{Tr} \rho^2(1-\rho)^2},
\end{eqnarray}
where $\kappa$ is the pairing tensor and potentials are given as:
\begin{equation}\label{pot1}
\Gamma^{\prime}_{\alpha \alpha^{\prime}} = 
\sum_{\beta \beta^{\prime}}V_{\alpha \beta
\alpha^{\prime} \beta^{\prime}}(\rho(1-\rho))_{\beta^{\prime} \beta}, 
\end{equation}
and 
\begin{equation}\label{pot2}
\Delta^{\prime}_{\alpha \beta} = \frac{1}{2} \sum_{\alpha^{\prime}
\beta^{\prime}}V_{\alpha \beta \alpha^{\prime} \beta^{\prime}}(\rho
\kappa)_{\alpha^{\prime} \beta^{\prime}},
\end{equation}
which can be calculated in a full analogy to $\Gamma$ and $\Delta$ by 
replacing $\rho$ and $\kappa$ by $\rho(1-\rho)$ and $\rho\kappa$, respectively. 
In case of the seniority-pairing interaction with strength $G$,
equation~(\ref{ll2}) can be simplified to
\begin{equation}\label{ll3}
\lambda_{2}=\frac{G}{4} \frac {{\rm Tr} (1-\rho)\kappa~ {\rm Tr} \rho \kappa  
- 2~{\rm Tr} (1-\rho)^2 \rho^2} 
{\left[{\rm Tr}\rho (1-\rho )\right]^{2}-2~{\rm Tr}\rho^{2}(1-\rho)^{2}}.
\end{equation}
%
The equation~(\ref{ll2}) can be well approximated by the seniority-pairing
expression~(\ref{ll3}) with the effective strength (G) and can be written 
in terms of pairing energy ($E_{\text{pair}}$) and average pairing gap
($\bar{\Delta}$)~\cite{ring05}:
\begin{equation}
G=G_{\text{eff}} = -\frac{\bar{\Delta}^2}{E_{\text{pair}}}\,
\end{equation}
where, $E_{\text{pair}} = -\tfrac{1}{2}{\rm Tr}\Delta \kappa$ and 
$\bar{\Delta}  = \frac{ { \rm Tr}\Delta \rho}{{ \rm Tr}\rho} $.
 
Here, calculations have done using the density dependent delta pairing force 
with the pairing strength $V_0$ = -244.72 MeV ${fm}^3$, pairing cut-off 
energy 60.0 MeV and pairing window 60.0 MeV. 
These quantities have been fitted to reproduce the neutron pairing gap of 
$^{120}$Sn which is consistent with Ref.~\cite{bena05}.
Average pairing gap ($\bar{\Delta}$) is obtained from the level density. 
Thus, it varies from nucleus to nucleus depending on
the density distribution of nucleons. The results for pairing gap
($\triangle_n$, $\triangle_p$), effective strength ($G_n$, $G_p$) and 
pairing energy ($E_{pair}$) for Zr isotopes are given in Table~\ref{tab3}.

\section{Theoretical Framework for Relativistic Mean Field Model}\label{rmf}
The relativistic mean field (RMF) model 
\cite{gamb90,patra91,wal74,sero86,horo81,bogu77,price87} is very 
successful in recent years for both finite nuclei and infinite nuclear matter
from normal to super-normal conditions.
In the present calculations, we have used the RMF Lagrangian~\cite{gamb90} 
with the NL3 parameter set~\cite{lala97}, which is quite successful for 
both $\beta$-stable and drip-lines nuclei.
The Lagrangian contains the terms of interaction between mesons and nucleons 
and also self-interaction of isoscalar scalar {\it sigma} meson. 
The other mesons are isoscalar vector {\it omega} and isovector vector
{\it rho} mesons. The photon field $A_{\mu}$ is included to take care of the 
Coulombic interaction of protons. 
A definite set of coupled equations are obtained from the Lagrangian 
which are solved self-consistently in an axially deformed Harmonic Oscillator 
(HO) basis with $N_F=N_B=12$, Fermionic and Bosonic oscillator quanta, 
respectively. 
A detail study about choosing the HO basis is given in 
subsection~\ref{sec:basis}.
The relativistic Lagrangian density for a nucleon-meson many-body systems 
is written as:
\begin{eqnarray}
{\cal L}&=&\overline{\psi_{i}}\{i\gamma^{\mu} \partial_{\mu}-M\}\psi_{i}
+{\frac12}\partial^{\mu}\sigma\partial_{\mu}\sigma
-{\frac12}m_{\sigma}^{2}\sigma^{2}
-{\frac13}g_{2}\sigma^{3} 
-{\frac14}g_{3}\sigma^{4}
-g_{s}\overline{\psi_{i}}\psi_{i}\sigma
\nonumber\\
&& 
-{\frac14}\Omega^{\mu\nu}
\Omega_{\mu\nu}
+{\frac12}m_{w}^{2}V^{\mu}V_{\mu}
-g_{w}\overline\psi_{i}
\gamma^{\mu}\psi_{i} V_{\mu}
-{\frac14}\vec{B}^{\mu\nu}.\vec{B}_{\mu\nu}
+{\frac12}m_{\rho}^{2}{\vec R^{\mu}}.{\vec{R}_{\mu}}
\nonumber\\
&&
-g_{\rho}\overline\psi_{i}\gamma^{\mu}\vec{\tau}\psi_{i}.\vec {R^{\mu}}
-{\frac14}F^{\mu\nu}F_{\mu\nu}-e\overline\psi_{i}
\gamma^{\mu}\frac{\left(1-\tau_{3i}\right)}{2}\psi_{i}A_{\mu}.
\label{lagrangian} 
\end{eqnarray}
Here, sigma meson field is denoted by $\sigma$, omega meson field by $V_{\mu}$
and rho meson field by ${\vec R_{u}}$ and $A_{\mu}$ denotes the
electromagnetic field, which couples to the protons. 
The Dirac spinors are
given by ${\psi}$ for the nucleons, whose third component of isospin is
denoted by $\tau_{3}$ and $g_{s}$, $g_2$, $g_3$, $g_{\omega}$, 
$g_{\rho}$ are the coupling constants.
The center of mass (c.m.) motion energy correction is estimated by the 
harmonic oscillator approximation $E_{c.m.} = \frac{3}{4}(41A^{-1/3})$. 
From the resulting proton and neutron quadrupole moments, the quadrupole 
deformation parameter  $\beta_{2}$ is defined as:
\begin{equation}  
Q = Q_n + Q_p = \sqrt{\frac{16\pi}5} \left(\frac3{4\pi} AR_0^2\beta_2\right),
\end{equation}
with $R_0=1.2 A^{1/3}$ (fm), and the root mean square matter radius 
are given as:
\begin{equation}
\langle r_m^2 \rangle = {1\over{A}}\int\rho(r_{\perp},z) r^2d\tau,
\end{equation}
where $A$ is the mass number, and $\rho(r_{\perp},z)$ is the deformed
density. The total binding energy and other observables are also obtained
by using the standard relations\cite{gamb90,patra91}.

\subsection{Pairing Correlations in RMF formalism}\label{sec:bcs}

The pairing correlation plays an important role in open shell nuclei to 
describe the ground state properties, like binding energy, charge radius, 
single particle energy level and deformation.
The relativistic Lagrangian contains only terms like $\psi^{\dag}\psi$,
and no terms of the form  $\psi^{\dag}\psi^{\dag}$ at the mean level.  
The inclusion of the pairing correlation of the form $\psi \psi$ and 
two-body interaction $\psi^{\dag} \psi^{\dag} \psi \psi$ in the Lagrangian 
violate the particle number conservation~\cite{patra93}. 
We used the pairing correlation externally in the RMF model. 
In our calculation, the constant gap BCS-approach take care the pairing 
correlation for open shell nuclei.
The general expression for pairing energy in terms of occupation 
probabilities $v_i^2$ and $u_i^2=1-v_i^2$ is written as~\cite{pres82,patra93}:
\begin{equation}
E_{pair}=-G\left[\sum_{i>0}u_{i}v_{i}\right]^2,
\end{equation}
with $G=$ pairing force constant. 
The variational approach with respect to $v_i^2$ gives the BCS equation 
\cite{pres82}:
\begin{equation}
2\epsilon_iu_iv_i-\triangle(u_i^2-v_i^2)=0,
\label{eqn:bcs}
\end{equation}
using $\triangle=G\sum_{i>0}u_{i}v_{i}$. 

The occupation number is defined as:
\begin{equation}
n_i=v_i^2=\frac{1}{2}\left[1-\frac{\epsilon_i-\lambda}{\sqrt{(\epsilon_i
-\lambda)^2+\triangle^2}}\right].
\end{equation}
The values of $\triangle$ for the nucleons (neutron and proton) is taken from 
the phenomenological formulae of Madland and Nix \cite{madland}:
\begin{eqnarray}
\triangle_n=\frac{r}{N^{1/3}}exp(-sI-tI^{2}),\qquad
\triangle_p=\frac{r}{Z^{1/3}}exp(sI-tI^{2}),
\end{eqnarray}
where, $I=(N-Z)/A$, $r=5.73$ MeV, $s=0.117$, and $t=7.96$.
 
The chemical potentials $\lambda_n$ and $\lambda_p$ are determined by the
particle numbers for neutrons and protons. 
Finally, the pairing energy is computed as:
\begin{equation}
E_{pair}=-\triangle\sum_{i>0}u_{i}v_{i}.
\end{equation}
For a particular value of $\triangle$ and $G$, the pairing energy $E_{pair}$
diverges, if it is extended to an infinite configuration space.
In fact, in all realistic calculations with finite range forces,
the contribution of states of large momenta above the Fermi surface
(for a particular nucleus) to  $\triangle$ decreases with energy.
Therefore, we use a pairing window, where the equations are extended up to the 
level $|\epsilon_i-\lambda|\leq 2(41A^{-1/3})$ which is the function of single 
particle energy. 
The factor 2 has been determined so as to reproduce the pairing correlation 
energy for neutrons in $^{118}$Sn using Gogny force 
\cite{sero86,patra93,dech80}. 
It is to be noted that recently Karatzikos et al.~\cite{karatzikos10} has shown
that if one use the constant pairing window which is adjusted for one state
at particular deformation then it may lead to errors at different energy 
solution (different state solution). 
However, we have not taken this problem into account in our calculations, as
we have adjusted to reproduce the pairing as a whole of $^{118}Sn$ nucleus.

\subsection{Selection of Basis Space}\label{sec:basis}
\begin{figure}[ht]
\includegraphics[width=1.\columnwidth]{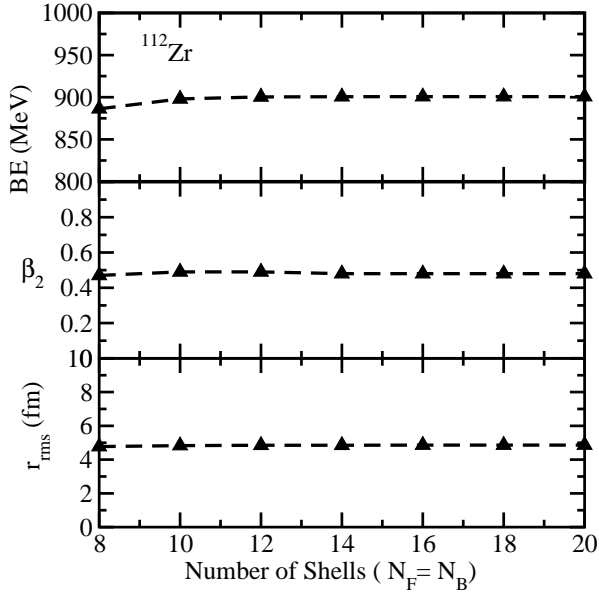}
\caption{(color online) The binding energy (BE), root mean square matter 
radius ($r_{rms}$) and quadrupole deformation parameter ($\beta_{2}$) with 
the harmonic oscillator basis.
}
\label{basis}
\end{figure}

After getting the self-consistent mean field equations for both Fermions and 
Bosons, we need to solve these equations by expanding the wave functions 
(potentials) in the deformed harmonic oscillator basis and solve the 
self consistent equations iteratively. 
For the exotic (drip-line) nuclei more harmonic 
oscillator quanta require to get the proper convergence of the system.
In our calculations, we used the harmonic oscillator quanta $N_F$=$N_B$=12,
where $N_F$ for Fermionic and $N_B$ for Bosonic quanta.
The convergence of the physical observable like binding energy (BE), root mean 
square matter radius ($r_{rms}$) and quadrupole deformation parameter 
($\beta_{2}$) with the harmonic oscillator basis are tested and obtained 
results are shown in Figure~\ref{basis}. 
We used the non-constraint calculation (free solution) to get the physical 
observables and for the large initial deformation basis parameter 
($\beta_{0}=0.6$), because we want to check the convergence for the 
large deformation basis parameter. 
If we increase the basis quanta from 12 to 14 the increment in the energy 
is $\sim 0.21$MeV which is near the accuracy of the theoretical models and 
by increasing the basis space the convergence time increases dramatically, 
so we use the optimum basis space which is suitable for the present 
calculations.
Thus $N_F$=$N_B$ $\ge$12 is enough for the convergence of the system which is 
shown in Figure~\ref{basis}.

To study the convergence of solutions in both RMF (NL3) and SHF (Sly4) 
formalisms, we have calculated the binding energy and corresponding quadrupole 
moment with different initial guess for the quadrupole moments. 
It is found that the calculated quadrupole deformation parameter 
$\beta_2$  is independent of the initial guess value of deformation $\beta_0$.
Both the formalisms give almost similar results except spherical solution 
obtained with an initial deformation $\beta_0$ = 0.03. Due to this 
suspicious behavior of the SHF(SLy4) result at the spherical solution, 
we ignore it for further analysis. We perform the free calculation for
 $^{82,100,102,104}Zr$ isotopes and calculated results are given in 
Table~\ref{tab2}. It is to be noted that from the potential energy surface curve 
as well as from the analysis of basis deformation, we get a spherical solution
for lighter isotopes of Zr, such as $^{82-92}Zr$. However, the zero 
deformation does not stable for heavier masses of Zr upto A = 106.
Again the appearence of zero solution get stabilize with increase mass number
(see PES curve).

\section{Calculations and Results}\label{result}

We used the non-constraint calculation in both the RMF and SHF formalisms.
For this, first we put some initial guess value of basis deformation parameter 
and let the system goes to find out the minimum energy state in local region 
corresponding to the initial guess i.e. $\beta_{0}$. 
We put the three guess values for each nucleus ($\beta_{0}=0.001, \pm0.3$).
In this case, final state (shape) of the nucleus may be different from the 
initial guess $\beta_{0}$ parameter.   
Both the SHF and RMF formalisms predict very good binding energy, 
root mean square (rms) radius and quadrupole deformation parameter $\beta_2$, 
not only for nuclei in stability line, but also for drip-lines nuclei. 
In this work, we have  analyzed the structure of proton and neutron-rich Zr 
nuclei and studied two important phenomena such as 
(i) shape co-existence and  
(ii) parity doublet for some specific Zr isotopes. 
For this, we obtain matter radius $r_m$, quadrupole deformation parameter 
$\beta_2$ and ground state binding energy from proton to neutron drip-lines.  
The calculated results are given in Table~\ref{tab1} and the  shape 
co-existence and parity doublets are shown in Figs.~\ref{bea}~to~\ref{100}.

\subsection{Potential Energy Surface (PES)}
\begin{figure}[ht]
\includegraphics[width=1.\columnwidth]{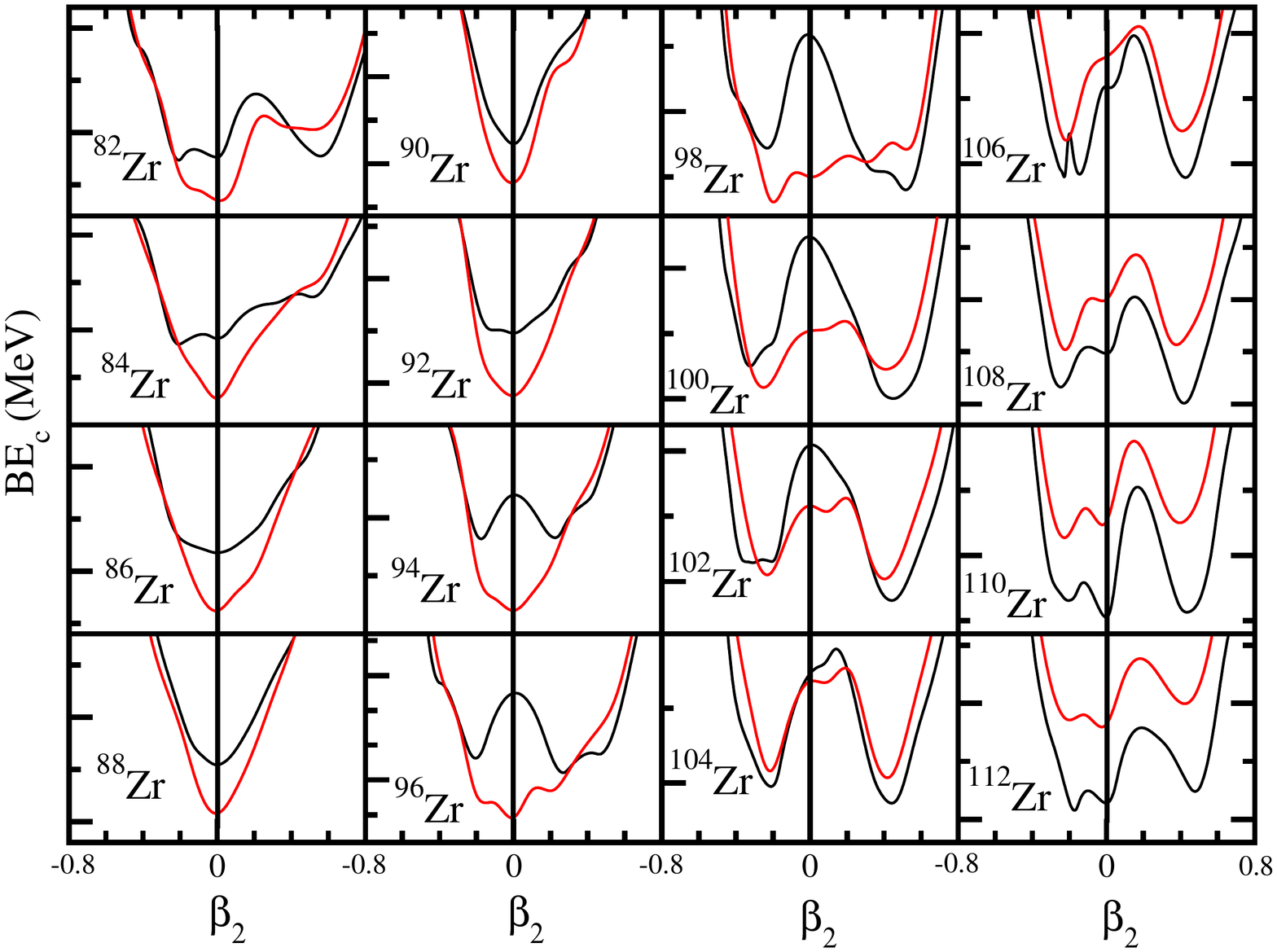}
\caption{\label{pes} (color online) The potential energy surface for the 
Zr isotopes for NL3 (black line) and SLy4 (red line) force parameter.
}
\end{figure}

In our considered Zr isotopes, many nuclei are deformed in their ground state
and for calculating the ground state properties one should include the 
deformation into the formalism. It may possible that some nuclei have 
almost same energy with different shape configurations (spherical, prolate 
or oblate), this type of states are known as shape coexistence. 
To get the solution with different deformations one should perform the 
constraint calculation as a function of quadrupole deformation parameter with 
various constraint binding energy ($BE^{c}$).
For constraint calculation, we minimized $<H^{'}>$ instead of $<H>$ 
which are related to each other by the following 
relation~\cite{patra09,flocard73,koepf88,fink89,hirata93}:

\begin{eqnarray}
H^{'}=H-\lambda Q,\qquad {with} \qquad Q=r^{2}Y_{20}(\theta, \phi),
\end{eqnarray}
where, $\lambda$ is the Lagrange multiplier which is fixed by the constraint
$<Q>_\lambda$ = $Q_{0}$. 
We have done the constraint calculation for Zr isotopes for both 
the parameter sets (NL3 and SLy4) and obtained results are shown in 
Figure~\ref{pes}. We get almost similar results in both the formalisms. 
For example, the three minima of $^{110}Zr$ are located at 
$\beta_2$ = -0.217, 0.0 and 0.398 respectively. Similar situation 
can be found for $^{108,112}Zr$ nuclei. The ground-state potential 
energy surfaces allow us to determine the equilibrium shapes 
(the lowest minimum). It is worthy to mention here that the 
minima near zero is not well developed, but may be considered 
as an isomeric state.

The $\gamma-$soft configuration~\cite{dobaczewski77} is (energy almost 
constant for a large range 
of deformation) obtained in both models which clearly noticed in 
Fig.~\ref{pes}.
For example, $^{84}$Zr isotopes have $\gamma-$soft surface near the 
$\beta_{2} \approx$-0.2 to 0.5 deformation in RMF (NL3) model and 
$^{98}$Zr also followed the same trends in SHF(SLy4) formalism. 
In these cases, a triaxial calculation is most welcome to get a detail account
on the potential  energy surfaces. 
The free and constraint solution are giving the similar results for the ground 
state energy and shape configuration.

\subsection{Binding energy and shape co-existence}
The nuclear binding energy (BE) is a physical quantity, which is precisely 
measured experimentally and is responsible for nuclear stability and 
structure of nuclei. The maximum binding energy corresponds to the ground 
state and all other solutions are intrinsic excited states of a nucleus.
These are not necessarily the lowest excitations, there could be 
rotational excitations below their first excited state, 
which is beyond the scope of our present calculations
for further analysis. 
The BE for Zr isotopes obtained by SHF(SLy4) and RMF(NL3) calculations 
are depicted in Table~\ref{tab1} and the results compare with experimental 
data~\cite{raman01, wang12, angeli13}, wherever available. 
From the ground  and excited 
intrinsic states binding energies, we have measured their difference 
$\triangle{BE}=BE(gs)-BE(es)$ and examined the shape co-existence 
phenomena.  When we find a small value of  $\triangle{BE}$, then we termed
it as a case of shape co-existence (degenerate solutions with different 
quadrupole deformations).
The shape co-existence means, there
is a maximum possibility of the nucleus, find in  either shapes.

\begin{table}
\hspace{2.0cm}
\caption{The binding energy BE (MeV), root mean square radii (fm), 
quadrupole deformation parameter $\beta_2$ for Zr isotopes. The
experimental results \cite{wang12,angeli13,raman01} are given for comparison.}
\renewcommand{\tabcolsep}{0.09cm}
\renewcommand{\arraystretch}{0.8}
{\begin{tabular}{|c|c|c|c|c|c|c|c|c|c|c|c|c|c|c|c|c|c|c|c|c|}
\hline
\hline
& \multicolumn{6}{ |c| }{RMF (NL3)} &\multicolumn{6}{ |c| }{SHF (SLy4)}& \multicolumn{3}{ |c| }{Expt.}  \\
\cline{2-16}
																														
Nucleus	&	$r_{ch}$	&	$r_n$	&	$r_p$	&	$r_{rms}$	&	        BE	&	$\beta_2$	&	   $r_{ch}$	&	     $r_n$	&	      $r_p$	&	 $r_{rms}$	&	      BE	&	  $\beta_2$	&	   $r_{ch}$	&	  $\beta_2$	&	      BE	\\
\hline	&	4.282	&	4.123	&	4.207	&	4.165	&	664.5	&	-0.172	&	4.276	&	4.125	&	4.201	&	4.163	&	665.9	&	-0.154	&		&		&		\\
$^{80}$Zr	&	4.274	&	4.112	&	4.198	&	4.155	&	665.9	&	0.000	&	4.262	&	4.108	&	4.186	&	4.147	&	669.2	&	0.000	&		&		&	669.9	\\
	&	4.362	&	4.207	&	4.288	&	4.248	&	665.1	&	0.480	&	4.393	&	4.243	&	4.319	&	4.281	&	665.3	&	0.5001	&		&		&		\\
	&		&		&		&		&		&		&		&		&		&		&		&		&		&		&		\\
	&	4.284	&	4.173	&	4.209	&	4.191	&	690.9	&	-0.191	&	4.284	&	4.163	&	4.208	&	4.185	&	691.8	&	-0.146	&		&		&		\\
$^{82}$Zr	&	4.272	&	4.158	&	4.197	&	4.177	&	691.7	&	0.000	&	4.269	&	4.151	&	4.193	&	4.172	&	694.4	&	0.000	&		&	0.367	&	694.5	\\
	&	4.371	&	4.262	&	4.297	&	4.279	&	689.6	&	0.480	&	4.372	&	4.248	&	4.298	&	4.272	&	690.4	&	0.430	&		&		&		\\
	&		&		&		&		&		&		&		&		&		&		&		&		&		&		&		\\
	&	4.324	&	4.453	&	4.249	&	4.367	&	809.8	&	-0.148	&	4.348	&	4.397	&	4.274	&	4.345	&	813.5	&	-0.133	&		&		&		\\
$^{94}$Zr	&		&		&		&		&		&		&	4.327	&	4.385	&	4.253	&	4.329	&	814.1	&	0.000	&	4.332	&	0.094	&	814.7	\\
	&	4.326	&	4.452	&	4.251	&	4.368	&	809.4	&	0.163	&	4.383	&	4.421	&	4.309	&	4.373	&	813.5	&	0.242	&		&		&		\\
		&&&&&&&&&&&		&		&		&		&		\\
	&4.355	&4.521	&4.281	&4.423	&822.9	&-0.191	&4.371	&4.449	&4.297	&4.386	&825.8	&-0.152	&&	&		\\
$^{96}$Zr	&	4.368	&	4.520	&	4.294	&	4.427	&	823.2	&	0.240	&	4.446	&	4.499	&	4.374	&	4.447	&	824.2	&	0.345	&	4.351	&	0.08	&	829.0	\\
	&		&		&		&		&		&		&		&		&		&		&		&		&		&		&		\\
	&		&		&		&		&		&		&		&		&		&		&		&		&		&		&		\\
		& 4.381	& 4.585	& 4.307	& 4.473	& 835.4	& -0.215 & 4.403 & 4.506 & 4.330& 4.435	& 838.4	& -0.196& & & \\
$^{98}$Zr	& 4.501	&4.676	&4.429	&4.577	&836.0	&0.497	&4.502	&4.579	&4.431	&4.519	&836.9	&0.430	& 4.401	& & 840.9 \\
		& 	&	&	&	&	&	&	&	&	&	&	&	&&&	\\
	 	&&&&&&&&&&&&&&		&		\\
	&	4.400	&	4.639	&	4.327	&	4.517	&	846.9	&	-0.217	&	4.427	&	4.555	&	4.354	&	4.476	&	849.4	&	-0.210	&		&		&		\\
$^{100}$Zr	&	4.487	&	4.690	&	4.415	&	4.582	&	847.7	&	0.445	&	4.512	&	4.612	&	4.441	&	4.545	&	849.7	&	0.421	&	4.49	&	0.355	&	852.2	\\
	&		&		&		&		&		&		&		&		&		&		&		&		&		&		&		\\
	&		&		&		&		&		&		&		&		&		&		&		&		&		&		&		\\
	&	4.416	&	4.687	&	4.343	&	4.555	&	858.0	&	-0.206	&	4.449	&	4.599	&	4.376	&	4.513	&	859.7	&	-0.215	&		&		&		\\
$^{102}$Zr	&	4.496	&	4.732	&	4.424	&	4.614	&	858.3	&	0.430	&	4.536	&	4.654	&	4.465	&	4.581	&	860.5	&	0.429	&	4.53	&	0.427	&	863.6	\\
	&		&		&		&		&		&		&		&		&		&		&		&		&		&		&		\\
	&		&		&		&		&		&		&		&		&		&		&		&		&		&		&		\\
	&	4.436	&	4.735	&	4.363	&	4.595	&	868.7	&	-0.207	&	4.469	&	4.64	&	4.397	&	4.548	&	869.6	&	-0.219	&		&		&		\\
$^{104}$Zr	&	4.404	&	4.715	&	4.331	&	4.571	&	865.2	&	0.000	&	4.426	&	4.617	&	4.353	&	4.517	&	866.8	&	0.000	&		&		&	873.8	\\
	&	4.512	&	4.780	&	4.441	&	4.652	&	867.9	&	0.424	&	4.557	&	4.696	&	4.486	&	4.617	&	870.2	&	0.430	&		&		&		\\
	&		&		&		&		&		&		&		&		&		&		&		&		&		&		&		\\
	&	4.461	&	4.783	&	4.389	&	4.639	&	878.5	&	-0.226	&	4.49	&	4.68	&	4.418	&	4.583	&	879.2	&	-0.223	&		&		&		\\
$^{106}$Zr	&	4.422	&	4.764	&	4.349	&	4.612	&	876.0	&	0.000	&	4.444	&	4.658	&	4.372	&	4.552	&	876.1	&	0.000	&		&		&	883.2	\\
	&	4.534	&	4.825	&	4.463	&	4.691	&	877.6	&	0.420	&	4.574	&	4.734	&	4.503	&	4.648	&	879.3	&	0.421	&		&		&		\\
	&		&		&		&		&		&		&		&		&		&		&		&		&		&		&		\\
	&	4.483	&	4.829	&	4.411	&	4.679	&	886.8	&	-0.232	&	4.509	&	4.717	&	4.437	&	4.616	&	887.5	&	-0.226	&		&		&		\\
$^{108}$Zr	&	4.439	&	4.816	&	4.366	&	4.654	&	886.5	&	0.000	&	4.463	&	4.697	&	4.391	&	4.586	&	885.2	&	0.000	&		&		&	891.7	\\
	&	4.556	&	4.873	&	4.485	&	4.733	&	886.8	&	0.420	&	4.592	&	4.769	&	4.522	&	4.679	&	887.6	&	0.414	&		&		&		\\
	&		&		&		&		&		&		&		&		&		&		&		&		&		&		&		\\
	&	4.487	&	4.865	&	4.415	&	4.707	&	894.3	&	-0.190	&	4.521	&	4.749	&	4.45	&	4.642	&	894.3	&	-0.210	&		&		&		\\
$^{110}$Zr	&	4.454	&	4.864	&	4.381	&	4.964	&	896.2	&	0.000	&	4.48	&	4.733	&	4.408	&	4.617	&	893.9	&	0.000	&		&		&	899.5	\\
	&	4.595	&	4.955	&	4.525	&	4.803	&	893.9	&	0.473	&	4.621	&	4.824	&	4.551	&	4.727	&	894.4	&	0.437	&		&		&		\\
	&		&		&		&		&		&		&		&		&		&		&		&		&		&		&		\\
	&	4.500	&	4.904	&	4.429	&	4.740	&	901.9	&	-0.171	&	4.53	&	4.775	&	4.459	&	4.664	&	901.1	&	-0.176	&		&		&		\\
$^{112}$Zr	&	4.469	&	4.903	&	4.397	&	4.729	&	902.7	&	0.000	&	4.496	&	4.765	&	4.424	&	4.646	&	901.1	&	0.000	&		&		&	906.5	\\
	&	4.620	&	5.001	&	4.550	&	4.845	&	900.5	&	0.480	&	4.645	&	4.874	&	4.576	&	4.77	&	900.9	&	0.453	&		&		&		\\
\hline
\hline
\end{tabular}\label{tab1} }
\end{table}
\begin{table}
\hspace{0.1cm}
\caption{The binding energy BE (MeV), quadrupole deformation parameter $\beta_2$ and basis deformation parameter $\beta_0$ for Zr isotopes.}
\renewcommand{\tabcolsep}{0.1cm}
\renewcommand{\arraystretch}{1.0}
{\begin{tabular}{|c|c|c|c|c|c|c|c|c|c|c|c|c|c|c|c|c|c|c|c|c|c|}
\hline
\hline
& \multicolumn{2}{ |c| }{RMF (NL3)} &\multicolumn{3}{ |c| }{SHF (SLy4)}  && \multicolumn{2}{ |c| }{RMF (NL3)} &\multicolumn{3}{ |c| }{SHF (SLy4)} \\  
\cline{2-6} \cline{8-12}											
Nucleus	&	BE	&	$\beta_2$	&	BE	&	$\beta_2$	&	$\beta_0$	&	Nucleus	&	BE	&	$\beta_2$	&	BE	&	$\beta_2$	&	$\beta_0$	\\
\hline
	&	691.0	&	-0.194	&	691.8	&	-0.182	&	-0.6	&		&	858.1	&	-0.208	&	859.7	&	-0.216	&	-0.6	\\
	&	691.0	&	-0.192	&	691.8	&	-0.178	&	-0.5	&		&	858.1	&	-0.206	&	859.7	&	-0.216	&	-0.5	\\
	&	690.9	&	-0.191	&	691.8	&	-0.146	&	-0.4	&		&	858.0	&	-0.205	&	859.7	&	-0.215	&	-0.4	\\
	&	690.9	&	-0.191	&	691.8	&	-0.164	&	-0.3	&		&	858.0	&	-0.206	&	859.7	&	-0.215	&	-0.3	\\
	&	690.8	&	-0.190	&	692.2	&	-0.117	&	-0.2	&		&	858.0	&	-0.206	&	859.7	&	-0.215	&	-0.2	\\
	&	691.7	&	0.000	&	692.5	&	-0.103	&	-0.1	&		&	858.0	&	-0.207	&	859.7	&	-0.215	&	-0.1	\\
$^{82}Zr$	&	691.7	&	0.000	&	694.4	&	0.000	&	0.03	&	$^{102}Zr$	&	858.2	&	0.419	&	857.3	&	0.000	&	 0.0.3	\\
	&	691.7	&	0.000	&	690.0	&	0.461	&	0.1	&		&	858.3	&	0.426	&	860.5	&	0.429	&	0.1	\\
	&	691.7	&	0.000	&	690.0	&	0.477	&	0.2	&		&	858.3	&	0.429	&	860.5	&	0.428	&	0.2	\\
	&	689.4	&	0.493	&	690.0	&	0.493	&	0.3	&		&	858.2	&	0.429	&	860.6	&	0.428	&	0.3	\\
	&	689.4	&	0.481	&	690.0	&	0.496	&	0.4	&		&	858.2	&	0.429	&	860.6	&	0.428	&	0.4	\\
	&	689.4	&	0.473	&	690.0	&	0.484	&	0.5	&		&	858.3	&	0.430	&	860.5	&	0.429	&	0.5	\\
	&	689.6	&	0.480	&	690.4	&	0.430	&	0.6	&		&	858.1	&	0.428	&	860.5	&	0.429	&	0.6	\\
	&		&		&		&		&		&		&		&		&		&		&		\\
	&	846.9	&	-0.218	&	849.4	&	-0.212	&	-0.6	&		&	868.7	&	-0.208	&	869.6	&	-0.220	&	-0.6	\\
	&	846.9	&	-0.217	&	849.4	&	-0.211	&	-0.5	&		&	868.7	&	-0.206	&	869.6	&	-0.219	&	-0.5	\\
	&	846.9	&	-0.216	&	849.4	&	-0.211	&	-0.4	&		&	868.7	&	-0.206	&	869.6	&	-0.219	&	-0.4	\\
	&	846.9	&	-0.217	&	849.4	&	-0.210	&	-0.3	&		&	868.7	&	-0.207	&	869.6	&	-0.219	&	-0.3	\\
	&	846.8	&	-0.218	&	849.4	&	-0.210	&	-0.2	&		&	868.6	&	-0.207	&	869.7	&	-0.219	&	-0.2	\\
	&	846.8	&	-0.218	&	849.5	&	-0.210	&	-0.1	&		&	868.6	&	-0.208	&	869.7	&	-0.219	&	-0.1	\\
$^{100}Zr$	&	847.6	&	0.423	&	847.5	&	0.000	&	0.03	&	$^{104}Zr$	&	865.2	&	0.000	&	866.8	&	0.000	&	0.03	\\
	&	847.7	&	0.440	&	849.7	&	0.423	&	0.1	&		&	865.1	&	0.035	&	870.3	&	0.430	&	0.1	\\
	&	847.7	&	0.449	&	849.7	&	0.422	&	0.2	&		&	868.0	&	0.424	&	870.3	&	0.430	&	0.2	\\
	&	847.7	&	0.445	&	849.7	&	0.421	&	0.3	&		&	868.0	&	0.424	&	870.3	&	0.430	&	0.3	\\
	&	847.6	&	0.440	&	849.7	&	0.422	&	0.4	&		&	867.9	&	0.424	&	870.3	&	0.430	&	0.4	\\
	&	847.6	&	0.436	&	849.7	&	0.423	&	0.5	&		&	867.9	&	0.423	&	870.3	&	0.430	&	0.5	\\
	&	847.6	&	0.433	&	849.6	&	0.422	&	0.6	&		&	867.9	&	0.424	&	870.2	&	0.430	&	0.6	\\

\hline       
\hline
\end{tabular}\label{tab2} }
\end{table}

\begin{table}
\caption{The pairing gap, effective strenth and pairing energy for Zr isotopes.}
\begin{tabular}{|c|c|c|c|c|c|c|c|c|}
\hline
\hline
	& \multicolumn{3}{ |c| }{RMF (NL3)} &\multicolumn{5}{ |c| }{SHF (SLy4)} \\ 
\cline{2-9}						
Nucleus	& $\triangle_n$	&	$\triangle_p$	&	$E_{pair}$	&	$G_n$	&	$G_p$	&	$\triangle_n$	&	$\triangle_p$	&	$E_{pair}$	\\
\hline
$^{80}Zr$	&	1.673	&	1.673	&	18.995	&	-0.162	&	-0.158	&	0.158	&	0.143	&	4.578	\\
$^{82}Zr$	&	1.633	&	1.669	&	19.078	&	-0.149	&	-0.157	&	0.217	&	0.142	&	5.911	\\
$^{94}Zr$	&	1.242	&	1.422	&	15.188	&	-0.126	&	-0.145	&	0.169	&	0.124	&	4.485	\\
$^{96}Zr$	&	1.170	&	1.361	&	13.882	&	-0.125	&	-0.144	&	0.183	&	0.118	&	4.889	\\
$^{98}Zr$	&	1.100	&	1.300	&	11.999	&	-0.125	&	-0.141	&	0.194	&	0.135	&	5.44	\\
$^{100}Zr$	&	1.031	&	1.238	&	11.411	&	-0.121	&	-0.140	&	0.222	&	0.133	&	6.156	\\
$^{102}Zr$	&	0.966	&	1.176	&	10.767	&	-0.118	&	-0.138	&	0.251	&	0.131	&	6.695	\\
$^{104}Zr$	&	0.903	&	1.115	&	9.886	&	-0.116	&	-0.134	&	0.163	&	0.103	&	4.183	\\
$^{106}Zr$	&	0.844	&	1.056	&	8.840	&	-0.114	&	-0.133	&	0.158	&	0.095	&	3.948	\\
$^{108}Zr$	&	0.787	&	0.999	&	7.741	&	-0.112	&	-0.132	&	0.139	&	0.091	&	3.517	\\
$^{110}Zr$	&	0.735	&	0.944	&	6.971	&	-0.109	&	-0.133	&	0.143	&	0.084	&	3.468	\\
$^{112}Zr$	&	0.685	&	0.892	&	6.959	&	-0.106	&	-0.132	&	0.124	&	0.082	&	3.058	\\
\hline
\hline
\end{tabular}\label{tab3} 
\end{table}

\begin{figure}[ht]
\includegraphics[width=1.\columnwidth]{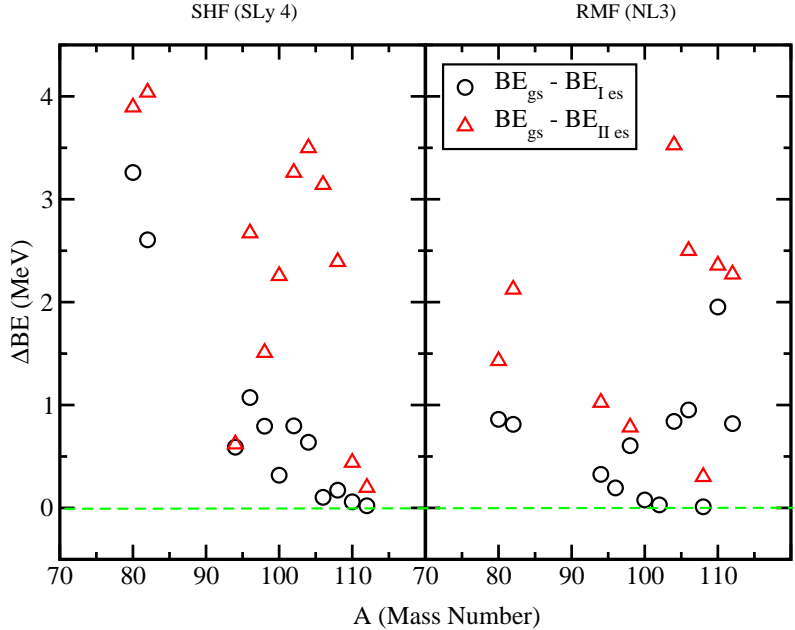}
\caption{(color online) The ground state binding energy difference from 
first and second intrinsic excited states for Zr isotopes. 
The zero reference point shown by the dashed horizontal line.
}
\label{bea}
\end{figure}

The binding energy difference between the ground and first and second
intrinsic excited states are shown in Fig.~\ref{bea} for Zr isotopes. 
The solid line
is the zero reference label, which marks the shape co-existence line. The
points which are on the line are designated as perfectly shape co-existence nuclei.  
The shape co-existence in $A=80$ mass region of nuclei using RMF formalism is
reported in Refs. \cite{patra93,rana,chakra13}. 
Here, it has shown that the
neutron deficient nuclei in this mass region possess spherical and 
largedeformed structures. In the present work, we would like to show that
not only the neutron-deficient Zr isotopes have shape co-existence, but also
other normal and neutron-rich Zr isotopes have low-lying largedeformed
configuration including the normal/spherical shape. Some times it so happens 
that the largedeformed solution becomes the ground state 
($^{98}Zr$, $\beta_2$=0.497 in RMF ) as shown in the Table~\ref{tab1}. The nuclei with 
shape co-existence shows the transition between the spherical to oblate 
to prolate due to minimum energy barrier between the shape co-existence states.
There are many isotopes ($^{96, 98, 100, 102, 108}$Zr), 
which have $\Delta BE \le 1$ MeV for both cases like 
$1^{st}$ and $2^{nd}$ intrinsic excited states. These type of shape
co-existence called triple shape co-existence \cite{petrovici12}.
If we see the $\Delta BE$ for $^{108}$Zr in  Figure~\ref{bea}, its excited 
state has almost same energy with its ground state, leading to the 
phenomenon of shape co-existence. These type of nuclei show the shape 
co-existence in their excited state and performed the shape change/ 
fluctuation in application of a small energy ($\le 1$ MeV). The shape 
co-existence is very important in the reaction study, because surface 
density distribution plays a crucial role in the cross-section and it will 
change by applying small perturbation in energy.  
Some isotopes of Zr are predicted to be triaxial ($\gamma \ne 0$) 
\cite{tomas11} in shape, which is one more
degree of freedom in shape orientation. The study of phenomenon is 
beyond the scope of this work, as we have used the axially symmetric 
formalism for the deformed nuclei.
 
Analyzing Fig.~\ref{bea} and the binding energy results of Table~\ref{tab1}, 
it is clear that the prediction of RMF(NL3) and SHF(SLy4) are almost 
similar. Again, comparing the results with experimental data, 
the SLy4 parameter set reproduce the data similar or even better than
NL3 set of the RMF formalism. In general, both the SHF and RMF have
tremendous predictive power upto a great extend of accuracy and can
be used the results to most part of the mass Table. 

\subsection{Evolution of single particle energy with deformation}

\begin{figure}[ht]
\includegraphics[width=1.\columnwidth]{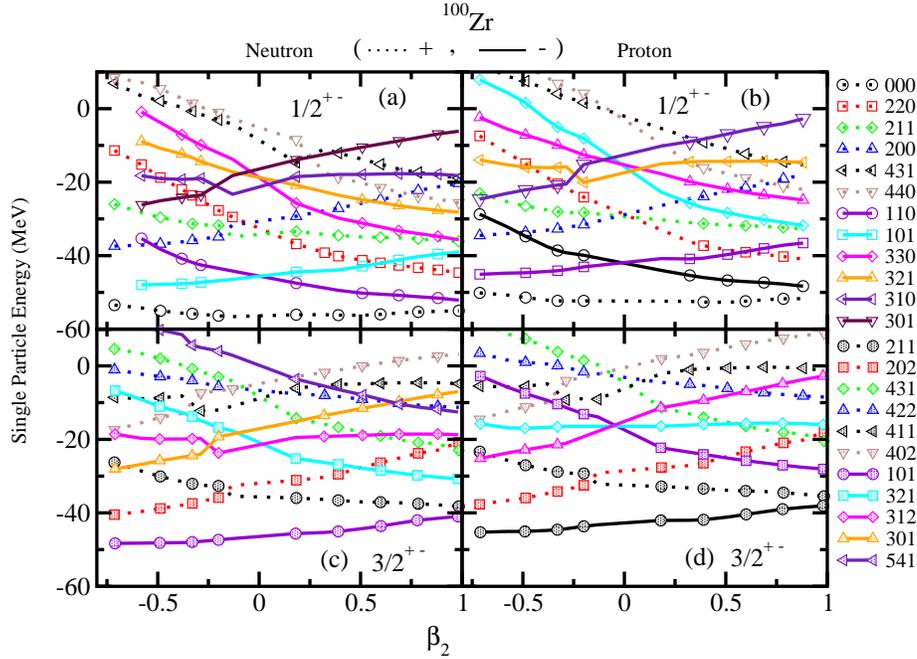}
\caption{\label{sp} (color online) Some selected single particle (s.p.) 
energy level evolution with deformation parameter $\beta_2$ for relativistic 
model by using NL3 parameter set. (a) 1/2+- s.p. levels for the neutron, 
(b) 1/2+- levels for proton, (c) 3/2+- s.p. levels for neutron, and 
(d) 3/2+- s.p. levels for proton. 
The positive parity (+) level is given by dotted line and 
negative parity levels (-) level is given by solid line.
}
\end{figure}

In this section, we have calculated the single particle energy of some selected 
Nilsson orbits with the different values of deformation parameter $\beta_2$ 
by the constraint calculation. 
The obtained results are given in Figure~\ref{sp}, where positive parity orbits 
shown by dotted and negative parity by solid lines for $^{100}$Zr isotope.
The single particle energy for neutron is given in Figure~\ref{sp}(a)
and proton single particle energy in Figure~\ref{sp}(b).
The lower level like ${\frac{1}{2}}^{+}[000]$ is very less affected by the 
variation of the deformation in both neutron and proton cases as shown in 
Figure~\ref{sp}(a,b). 
But as increase the energy of the levels, variation of single particle energy 
is also increases with the deformation parameter as shown in Figure.
We have plotted similar curve for ${\frac{3}{2}}^\pm$ orbits for the same 
nucleus $^{100}$Zr and obtained results are given in Figure~\ref{sp}(c,d) 
for neutron and proton, respectively. 
The evolution of single particle energy levels with 
deformation parameters followed similar nature of ${\frac{1}{2}}^{\pm}$ orbits. 
We repeated the calculation in non-relativistic SHF model also and 
obtained almost similar trend of levels, so we are not presenting the SHF 
results for the single particle energy evolution with deformation in the 
present manuscript. 
The single particle energies are evolved with the deformation parameter 
and opposite parity orbits are come closer with deformation.
A detail study is done in next section, where we will discuss about the 
parity doublets in the orbits at large deformation.

\subsection {Largedeformed configuration and parity doublet}
\begin{figure}[ht]
\includegraphics[width=1.0\columnwidth]{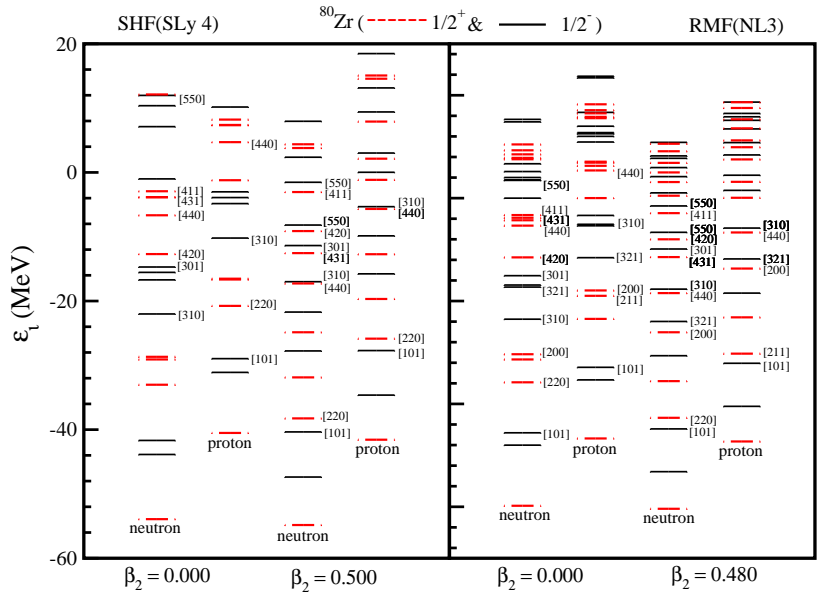}
\caption{ (color online) Single-particle levels for $^{80}$Zr in normal and
largedeformed states. 
The single-particle levels are denoted by the Nilsson indices 
[N,~$n_z$,~$\Lambda$]$\Omega^{\pi}$.}
\label{80}
\end{figure}

\begin{figure}[ht]
\includegraphics[width=1.0\columnwidth]{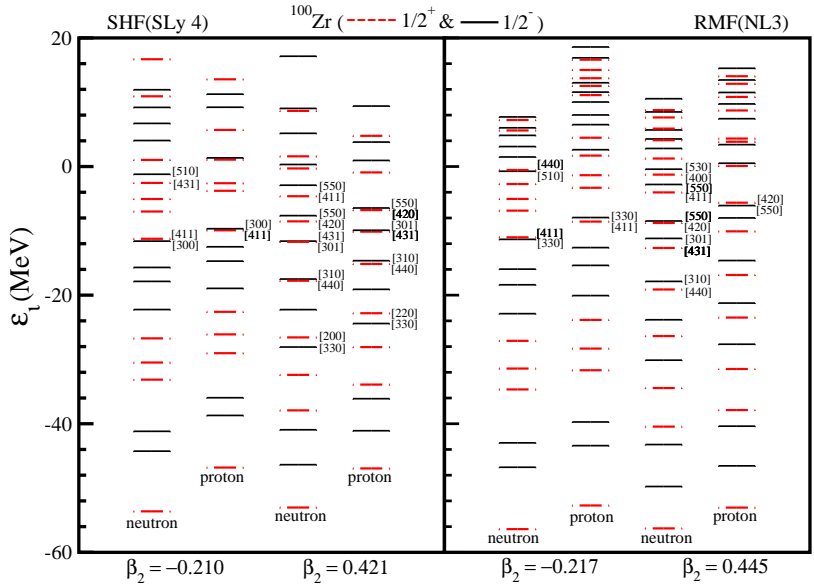}
\caption{ (color online) Single-particle levels for $^{100}$Zr isotopes 
in oblate and largedeformed states.}
\label{100}
\end{figure}

The parity doublet is an interesting configuration for the largedeformed
state of a nucleus. Recently, it is reported by Singh et al.
\cite{singh14} that, there exist a parity doublet in the largedeformed 
configuration for light mass nuclei. In the present calculations, we have
extended the investigation to relatively heavier mass region of the periodic
chart. In this case, we focused our study for Zr isotopes, where shape
co-existence is an usual phenomenon. In most of the cases of Zr isotopes, we
get a spherical or a normal deformed solution along with a largedeformed state
 both in the RMF(NL3) and SHF(SLy4) calculations. 
The evolution of single particle energy with deformation parameter $\beta_2$
for some selected nuclei are depicted in Figs.~\ref{80}~and~\ref{100}. 
The parity doublets are marked by their asymptotic quantum number 
[N,~$n_z$,~$\Lambda$], where N is principle quantum number, $n_z$ is number 
of nodes of the wave function in the $z-$direction (the number of times the 
radial wave function crosses zero). Larger $n_z$ values corresponds to wave 
function more extended in the $z-$direction which means lower energy orbits,
$\Lambda$ is the projection of the orbital angular momentum on to the $z-$axis.
Similar to the case of light mass nuclei \cite{singh14}, in case of Zr 
isotopes also,
the deformation-driving $\Omega^{\pi} = \frac{1}{2}^-$orbits come down in energy
in largedeformed solutions from the shell above, in contrast to
the normal deformed solutions. 
For each nucleus, we have compared the normal/spherical deformed
and the largedeformed configurations single particle energy orbits and 
analyzed the parity doublets states and some of them are given in this work.
The occurrence of approximate
$\frac{1}{2}^+$, $\frac{1}{2}^-$ parity doublets (degeneracy of $\Omega^{\pi}$=
$\frac{1}{2}^+$, $\frac{1}{2}^-$ states) for the largedeformed solutions
are clearly seen in Figs.~\ref{80}~and~\ref{100}, where excited largedeformed 
configurations for $^{80}$Zr and $^{100}$Zr are given.
As shown in Figure~\ref{80}, the energy level for spherical shape for 
opposite parity are well separated from  each other, but becomes closer 
with deformation which shows the parity doublets in the system.  
For example, in case of $^{80}$Zr, if we plot the single particle energy level for neutron, then the 
energy levels [310]$\frac{1}{2}^-$ and [440]$\frac{1}{2}^+$ are far from each 
other ($\sim$ 18.28 MeV in RMF), but becomes almost degenerate ($\sim$ 1.28 MeV) 
at largedeformation ($\beta_2=0.480$).
Same behavior we found in the single particle energy orbits 
[440]$\frac{1}{2}^+$ and [310]$\frac{1}{2}^-$ of proton intrinsic single 
particle energy distribution, i.e. in normal deformation, these two levels are
separated from each other by 16.8 MeV, but in largedeformed case 
($\beta_2$=0.480), it becomes closer ($\sim$ 0.5 MeV).
Qualitatively, the same behavior appears in the SHF(SLy4) results also
(left panel of the Figure). 
In Figure~\ref{100}, for $^{100}$Zr  we have given the 
largedeformed orbits for prolate and oblate cases both for RMF(NL3)
and SHF(SLy4) models. Here also, we are getting the 
parity doublet in oblate and prolate shapes, which implies that parity 
doublets are driving by the deformation and it will occur at the large 
deformation. 
Some parity doublet orbits are shown by Nilsson representation 
[N,~$n_z$,~$\Lambda$] in Figure~\ref{100}. 
If we put close inspection on Figure~\ref{100}, 
then in the oblate level of neutron, we  get several parity doublets like  
([411] [330]), ([440] [510]) and for proton ([330] [411]). 
For prolate case, the neutron parity doublet orbits ([530] [400]), ([550] [420]), 
([301] [431]), ([310] [440]) etc, similarly for the proton case. 

\section{Summary and Conclusions}\label{remark}
We calculate the ground and low-lying excited state properties,
like binding energy and quadrupole deformation parameter $\beta_2$ 
using RMF(NL3) and SHF(SLy4) formalisms for Zr isotopes near the drip-line regions.
In general both the RMF and SHF predict very good results throughout the isotropic
chain. We are getting the double and triple shape co-existence from 
our analysis in some Zr isotopes, which is consistent with the earlier 
data. The present prediction of parity doublet may be a challenge for 
the experimentalist to look for such configuration states. 
In general, we find large deformed solutions for the neutron-drip
nuclei, which agree with the experimental measurements.
In the calculations, a large number of low-lying intrinsic
largedeformed excited states are predicted in many of the isotopes, 
which shows the parity doublet near the Fermi levels. The parity 
doublet levels are nearly  degenerated in excited states which 
can make the two different parity band by transition of two particles 
from reference frame to these degenerate opposite parity levels.
It may be solved the problem of existence of the twin bands and 
quantization of alignments of shapes. 
This analysis will help us to understand the intrinsic excited states 
of the Zr and other similar isotopes. In this respect, some more 
calculations are required to build a general idea about the omega parity 
doublets.


\end{document}